\documentclass[journal=ancac3,manuscript=article,layout=traditional]{achemso}
\usepackage{amsmath}
\usepackage{xcolor}
\usepackage{graphicx}
\usepackage{dcolumn}
\usepackage{bm}
\usepackage{color}
\usepackage{ulem}
\usepackage{lipsum}
\usepackage{gensymb}
\usepackage{braket}
\usepackage[percent]{overpic}
\usepackage[version=4]{mhchem}

\title{MnRhBi$_3$: A Cleavable Antiferromagnetic Metal}

\author{Eleanor M. Clements}
\affiliation{Materials Science and Technology Division, Oak Ridge National Laboratory, Oak Ridge, Tennessee 37831, USA}
\alsoaffiliation{Neutron Scattering Division, Oak Ridge National Laboratory, Oak Ridge, TN 37831, USA}

\author{Dmitry Ovchinnikov}
\affiliation{Department of Physics and Astronomy, University of Kansas, Lawrence, KS 66045, USA
}

\author{Parul R. Raghuvanshi}
\affiliation{Materials Science and Technology Division, Oak Ridge National Laboratory, Oak Ridge, Tennessee 37831, USA}

\author{Valentino R. Cooper}
\affiliation{Materials Science and Technology Division, Oak Ridge National Laboratory, Oak Ridge, Tennessee 37831, USA}

\author{Satoshi Okamoto}
\affiliation{Materials Science and Technology Division, Oak Ridge National Laboratory, Oak Ridge, Tennessee 37831, USA}

\author{Andrew D. Christianson}
\affiliation{Materials Science and Technology Division, Oak Ridge National Laboratory, Oak Ridge, Tennessee 37831, USA}

\author{Joseph A. M. Paddison}
\affiliation{Neutron Scattering Division, Oak Ridge National Laboratory, Oak Ridge, TN 37831, USA}

\author{Brenden R. Ortiz}
\affiliation{Materials Science and Technology Division, Oak Ridge National Laboratory, Oak Ridge, Tennessee 37831, USA}

\author{Stuart Calder}
\affiliation{Neutron Scattering Division, Oak Ridge National Laboratory, Oak Ridge, TN 37831, USA}

\author{Andrew F. May}
\affiliation{Materials Science and Technology Division, Oak Ridge National Laboratory, Oak Ridge, Tennessee 37831, USA}

\author{Xiaodong Xu}
\affiliation{Department of Physics, University of Washington, Seattle, Washington 98195, United States}
\alsoaffiliation{Department of Materials Science and Engineering, University of Washington, Seattle, Washington 98195, United States
}

\author{Jiaqiang Yan}
\affiliation{Materials Science and Technology Division, Oak Ridge National Laboratory, Oak Ridge, Tennessee 37831, USA}

\author{Michael A. McGuire}
\affiliation{Materials Science and Technology Division, Oak Ridge National Laboratory, Oak Ridge, Tennessee 37831, USA}
\email{mcguirema@ornl.gov}

\newcommand{{\MnRhBi}}{MnRhBi${_3}$}

\newcommand{\TNI}{T$_{\rm N1}$}
\newcommand{\TNII}{T$_{\rm N2}$}

\newcommand{\qmagI}{$\textbf{q}_{\rm mag1}$}


\begin{document}


\begin{abstract}

Cleavable metallic antiferromagnets may be of use for low-dissipation spintronic devices; however, few are currently known. Here we present orthorhombic \MnRhBi\ as one such compound and present a thorough study of its physical properties. Exfoliation is demonstrated experimentally, and the cleavage energy and electronic structure are examined by density functional theory calculations. It is concluded that \MnRhBi\ is a van der Waals layered material that cleaves easily between neighboring Bi layers, and that the Bi atoms have lone pairs extending into the van der Waals gaps. A series of four phase transitions are observed below room temperature, and neutron diffraction shows that at least two of the transitions involve the formation of antiferromagnetic order. Anomalous thermal expansion points to a crystallographic phase transition and/or strong magnetoelastic coupling. This work reveals a complex phase evolution in \MnRhBi\ and establishes this cleavable antiferromagnetic metal as an interesting material for studying the interplay of structure, magnetism, and transport in the bulk and ultrathin limits as well as the role of lone pair electrons in interface chemistry and proximity effects in van der Waals heterostructures.

\end{abstract}

Since the initial demonstrations of magnetic order in the atomically thin limit in samples exfoliated from bulk crystals \cite{Lee2016, Wang2016, Gong2017, huang2017layer}, cleavable, van der Waals layered magnetic materials and their behaviors have emerged as important areas of research engaging the condensed matter physics, chemistry, and materials science communities \cite{duong2017van, burch2018magnetism, huang2020emergent, mcguire2020cleavable, wang2022magnetic}. The experimental virtue of cleavable magnetic materials is that single or few layer specimens can be obtained in a relatively straightforward way and investigated by themselves, incorporated into devices, and stacked with other materials to realize new properties and functionalities. \cite{geim2013van, lam2022morphotaxy, kim2023van}
The stacking degrees of freedom of single or few layers give rise to a plethora of exotic phenomena that are intensively investigated recently for their unique electronic, magnetic, and optical properties \cite{sun2024twisted, fox2023stacking, gong2019two}. This variability in stacking allows for the manipulation of the interlayer interactions that can dramatically affect physics and phenomena.

One of the areas that has been heavily influenced by the rise of cleavable magnetic materials is antiferromagnetic spintronics, which may provide new ways to realize the next generation of low dissipation devices \cite{Jungwirth2016,gomonay2018antiferromagnetic,Baltz2018, rahman2021recent}. Currently, much work in the field is focused on insulating examples \cite{liu2023recent, mcguire2017crystal, burch2018magnetism, wang2022magnetic, jiang2021recent}. However, metals may offer unique advantages with their electrical conductivity coupled with magnetism \cite{siddiqui2020metallic}. For example, the electrical conductivity in metals allows for the direct manipulation of spin currents through electric fields. Such spin manipulation, a fundamental aspect of spintronics, is more challenging in antiferromagnetic insulators.

Among the various 2D van der Waals antiferromagnets, few metallic systems have been identified. These include rare earth tritellurides, which have been extensively studied for their incommensurate charge density wave characteristics, diverse antiferromagnetic ground states, superconductivity under high pressure, and exceptionally high mobility.\cite{DiMasi1995,hamlin2009pressure,lei2020high, yumigeta2021advances} Others are found in the Fe$_n$GeTe$_2$ family, where the partial substitution of Fe with Co switches the materials from ferromagnetic to antiferromagnetic, \cite{may2019ferromagnetism, May2020,tian2020tunable,seo2021tunable} and CeSiI, which also brings heavy fermion physics into play.\cite{CeSiIneutron,CeSiInature} A challenge for the field is then to design and discover additional cleavable metallic antiferromagnets.

\begin{figure} \includegraphics [width = 3.5in] {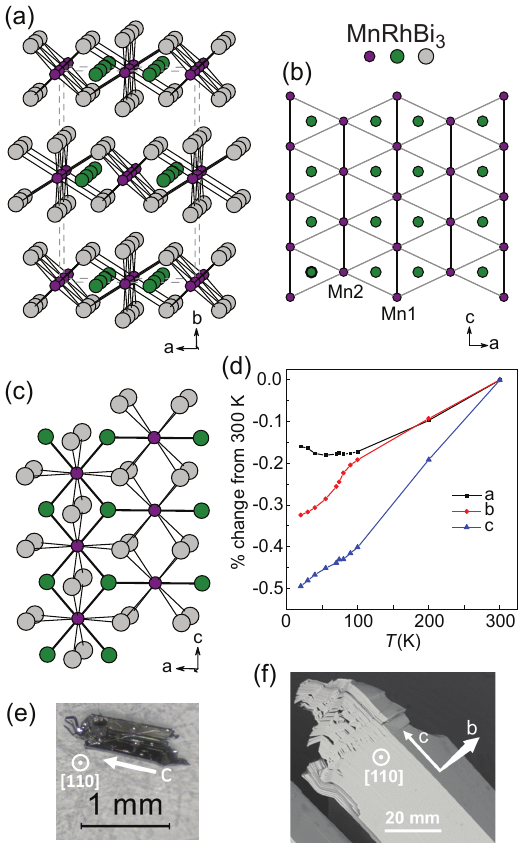}
\caption{
(a) The crystal structure of \MnRhBi\ highlighting the layered nature of the structure. The view is nearly along the \textit{c}-axis with the orthorhombic unit cell outlined by the dashed grey line. For clarity, only Mn-Bi bonds are shown.
(b) A single layer of composition MnRh viewed along the \textit{b}-axis, the stacking direction, and showing chains of Mn1 and chains of Mn2 running along the \textit{c}-axis. Short intrachain Mn-Mn distances are shown as black lines. Longer interchain Mn-Mn distances are shown as grey lines.
(c) A view along the \textit{b}-axis showing the coordination environments around the Mn sites with Mn-Rh and Mn-Bi bonds drawn. In this view the Mn chains run up the page.
(d) Temperature dependence of the lattice parameters of \MnRhBi\ relative to their values at 300\,K determined by powder x-ray diffraction.
(e) An as-grown crystal with orientation indicated.
(f) An SEM image of a crystal pressed onto carbon tape showing both the growth facet [110] and cleavage plane normal to the \textit{b} axis.
}
\label{structure}
\end{figure}

Here, a study of the magnetism, electrical transport, and cleavage of \MnRhBi\ is reported. Discovered in 2018, \MnRhBi\ is a relatively unexplored layered compound with van der Waals-like gaps between buckled Bi layers.\cite{kainzbauer2018single, kainzbauer2020reassessment} The structure is shown in Figure \ref{structure} and will be discussed in more detail below. In this work, a growth procedure for \MnRhBi\ crystals was developed, producing crystals with typical mass of 2-3 mg. Powder neutron diffraction and measurements of the thermodynamic, magnetic and transport properties in polycrystalline and single crystal samples show that \MnRhBi\ is a metallic antiferromagnet with a complex phase evolution below room temperature showing four transitions involving the magnetism and/or crystal structure.  Density functional theory (DFT) calculation of the cleavage energy and experimental exfoliation show that \MnRhBi\ crystals cleave easily, and electron localization function calculations show Bi lone pairs extend into the van der Waals gaps. With the interesting structural and physical properties presented here and the many promising avenues for further exploration, \MnRhBi\ provides an exciting new platform to study the interplay of structure, magnetism, and dimensionality in a metallic antiferromagnet.

\section{Results and Discussion}

\subsection{Crystal structure and cleavage of \MnRhBi}

Rietveld refinement of neutron powder diffraction (NPD) data, indexing of x-ray powder diffraction data, and single crystal x-ray diffraction data, show that the polycrystalline and single crystal \MnRhBi\ samples have the orthorhombic ($Cmmm$) crystal structure reported previously in Refs. \citenum{kainzbauer2018single, kainzbauer2020reassessment}. The lattice parameters from Rietveld refinement of powder x-ray diffraction data collected at room temperature are listed in Table \ref{table0} and are similar to those reported in Ref. \citenum{kainzbauer2020reassessment}. Single crystal diffraction revealed twinning that precluded high quality structural refinement. Therefore, the NPD-determined structure, refined using data collected at 100\,K, is used to discuss the structure here. The refined structural parameters are compiled in Table \ref{table0}, and the data and fit are shown in the Supporting Information. The lattice parameters show the expected thermal contraction when compared to the room temperature lattice parameters noted above.

\begin{table}
\caption{\label{table0} Lattice parameters of \MnRhBi\ at room temperature from powder x-ray diffraction, lattice parameters and atomic positions at 100\,K from neutron powder diffraction, and relaxed lattice parameters and atomic positions from density functional theory (DFT). The space group is $Cmmm$. Bi1 is at (0, y, 0), Bi2 is at (x, y, $\frac{1}{2}$), Mn1 is at ($\frac{1}{2}$, 0, $\frac{1}{2}$), Mn2 is at (0, 0, 0), and Rh1 is at (x, 0, 0). }
\begin{tabular}{cccc}
\hline
 & 293\,K & 100\,K & DFT\\
\hline
    a ({\AA}) & 8.849(1) & 8.8398(2) & 8.855 \\
    b ({\AA}) &  13.6739(7) & 13.6528(6) & 13.775 \\
    c ({\AA}) & 4.1295(5) & 4.11582(11) & 4.144 \\
    y-Bi1 & &  0.3377(4) & 0.3363 \\
    x-Bi2 & & 0.1943(3) & 0.1982 \\
    y-Bi2 & &  0.1244(2) & 0.1238 \\
    x-Rh1 & & 0.3012(6) & 0.3041 \\
\hline
\end{tabular}
\end{table}

Figures \ref{structure}a-c show the layered crystal structure of \MnRhBi. The layers are stacked along the crystallographic \textit{b}-axis. The structure of an individual layer is illustrated in Figure \ref{structure}b and \ref{structure}c. Each contains a flat Mn-Rh plane between the Bi layers. This plane contains a distorted triangular lattice of Mn with two inequivalent Mn sites, with chains of each type running along the \textit{c} direction. The intrachain Mn-Mn distance is 4.12 {\AA}, and the shortest interchain Mn-Mn distance is 4.88 {\AA}. The shortest interatomic distances within the structure are between Mn and Rh atoms, at 2.71 {\AA} ($\times$4) for Mn1-Rh and 2.66 {\AA} ($\times$2) for Mn2-Rh. This is close to the Mn-Rh distance found in binary intermetallics.\cite{Kouvel1963, kainzbauer2020reassessment} Figure \ref{structure}c shows the coordination around the Mn sites. The coordination environments are quite different, but both have high coordination numbers if Rh and Bi bonds are considered. Each Mn is surrounded by eight Bi atoms with none closer than 3.0 {\AA}. Rhodium is coordinated by three Mn atoms in the \textit{ac} plane at the distances noted above, and by six Bi atoms at 2.83 {\AA}, three above and three below the plane.

The anisotropy of the crystal structure is reflected in the growth habit of single crystals. Crystals grow in a needle-like morphology along the \textit{c}-axis, which is the direction of the Mn-Mn chains (see Figure \ref{structure}). In some cases the needles are grown together into a plate-like structure as shown in Figure \ref{structure}e. The flat surface of these plates was confirmed by x-ray diffraction to be the (110) plane. This is unusual given the crystal structure (the layers are stacked along the \textit{b}-axis), and the reason for this growth habit is unclear. Inspection of a crystal pressed into carbon tape for scanning electron microscopy better reveals the layered nature of the material. Figure \ref{structure}f shows the flat [110] growth facet, along with the cleavage planes perpendicular to the \textit{b}-axis.

The crystal structure suggests that the \MnRhBi\ should cleave between planes of Bi atoms. The closest contact between Bi atoms in neighboring slabs, across the van der Waals-like gap, is 3.57 {\AA}. In Bi metal, which is considered vdW layered, the interlayer distance across the gap is 3.53\,{\AA} at room temperature (3.44 {\AA} at 78\,K).\cite{Cucka1962} This suggests that \MnRhBi\ should cleave easily perpendicular to its \textit{b} axis. To examine this further, cleavage energy calculations were performed using DFT as described in the Methods section and the Supporting Information. The calculated energy for cleaving between the Bi layers is 0.56\,J/m$^2$. This is a relatively low value, and can be compared to calculated values for other van der Waals layered compounds, in the same units: graphite (0.43) \cite{Bjorkman-2012}, \ce{MoS2} (0.27)\cite{Bjorkman-2012}, \ce{CrGeTe3} and \ce{CrSiTe3} (0.35-0.38)\cite{Li-2014}, \ce{CrCl3} and \ce{CrI3} (0.3)\cite{McGuire-2015, ZhangCrX3-2015}, 113 or 226 type transition metal chalcophosphates (0.35$-$0.55) \cite{ZhangMPS3-2015}, and \ce{CrTe3} (0.5) \cite{CrTe3}. Thus, it is reasonable to consider \MnRhBi\ a van der Waals layered material.

Mechanical exfoliation of \MnRhBi\ was explored as described in the Methods section. Exfoliation on Si/\ce{SiO2} substrates produced flakes with lateral dimensions of tens of $\mu$m and thicknesses down to 20-30 nm that are suitable for device fabrication. These were used for electrical transport measurements discussed below. While the characterization of ultrathin flakes is ongoing and not the focus of the present work, an example of a 2\,nm thick sample exfoliated onto gold \cite{Velicky2018} is shown in the Supporting Information. The identification of \MnRhBi\ as a cleavable, van der Waals layered material based on its structure, growth habit, and calculated cleavage energy is experimentally confirmed by this exfoliation behavior.

\begin{figure} \centering \includegraphics [width = 3.5 in] {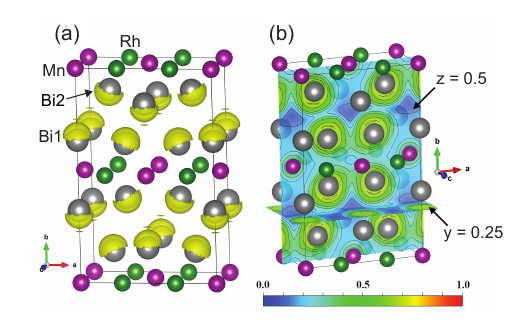}
\caption{Results of electron localization function (ELF) calculations. (a) Isosurfaces at ELF = 0.56 making the Bi lone pairs clearly visible. (b) Contour maps on the horizontal y = 0.25 plane passing through the vdW gap and the vertical z = 0.5 plane passing through Bi2 atoms and the closest Bi-Bi contact across the gap. Visualization of the ELF data was done using VESTA \cite{vesta}.
}
\label{elf}
\end{figure}

In most van der Waals layered compounds, including those with cleavage energies listed above, the van der Waals gap separates layers of closed-shell anions (divalent chalcogens or monovalent halides in those cases) that reside at the top and bottom surfaces of each structural slab. In the intermetallic compound \MnRhBi\ the situation is different, since the gap is between Bi layers and formal valences cannot easily be assigned. This is reminiscent of layered Zintl phases that have van der Waals gaps separating layers of Sn \cite{Arguilla2016, Asbrand1995, Eisenmann1991}.  In addition, Bi is known to have stereoactive lone pairs associated with asymmetric coordination environments in some materials \cite{laurita2022chemistry}. Since the van der Waals gap provides a one-sided coordination environment, it is natural to consider the relationship between this gap and Bi lone pairs. This was explored by electron localization function (ELF) calculations \cite{ELF1997}. The results are shown in Figure \ref{elf}, where ELF isosurfaces and planar contour plots are overlaid with the crystal structure. The asymmetric electron density around Bi is apparent. The regions of high density, associated with the Bi lone pairs, are indeed directed into the van der Waals gaps, which are at y = 0.25 and 0.75. The z = 0.5 plane contains the shortest cross-gap Bi-Bi contact in the structure, which is between atoms at the Bi2 positions. The contour plot at this plane clearly shows a minimum in electron density between the Bi2 atoms on each side of the gap, and the lone pairs are rotated slightly to avoid one another. Thus, \MnRhBi\ may be expected to have dangling lone pairs at its surface. It would be interesting to understand how this might affect the surface chemistry and physics of cleaved \MnRhBi\ crystals. The lone pairs may influence surface reconstruction, catalytic properties, and quantum mechanical proximity effects, all of which are worthy of further study.

The temperature evolution of the crystal structures gives information about phase transitions that may be crystallographically driven or couple strongly to the lattice. Lattice parameters of \MnRhBi\ determined from powder x-ray diffraction data collected between 20 and 300\,K are shown in Figure \ref{structure}d. While the room temperature structure described above accounts for the diffraction data well over the entire temperature range, anomalous temperature dependence is seen in all three lattice parameters upon cooling below about 80\,K. This indicates the presence of a phase transition that couples to the lattice, but not strongly enough to break the symmetry of the crystal structure within the sensitivity of our x-ray and neutron diffraction measurements.

\subsection{Neutron diffraction and magnetic order}

\begin{figure*} \centering \includegraphics [width = \textwidth] {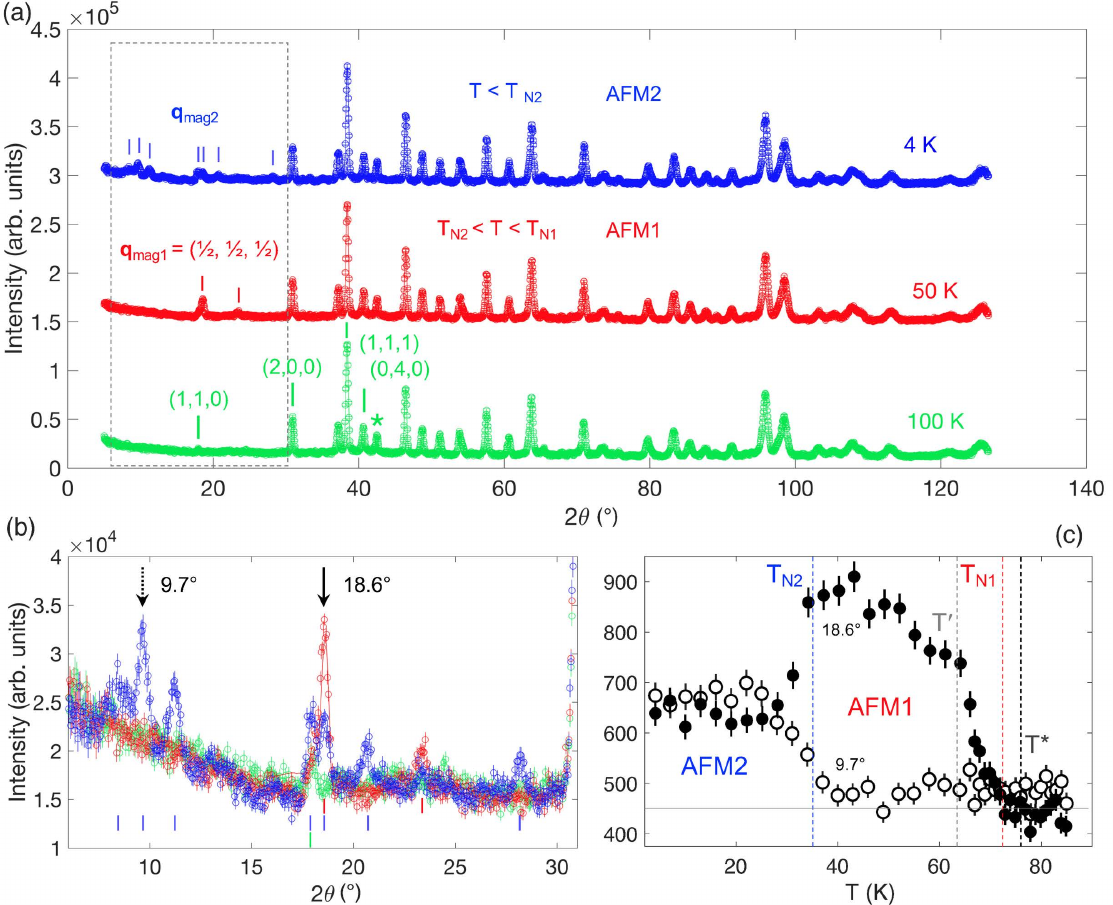}
\caption{
Neutron diffraction from powders ground from \MnRhBi\ crystals ($\lambda = 2.41$ \AA). (a) Diffraction patterns collected on HB-2A at 100, 50, and 4\,K. Ticks and labels with the 100\,K data identify several nuclear Bragg peaks, and the asterisk marks a peak from residual Bi metal from the crystal growth. Ticks with the 50 and 4\,K data mark magnetic Bragg peaks. At 50\,K, magnetic Bragg peaks are indexed with a reduced wave vector \qmagI $= (\frac{1}{2},\frac{1}{2},\frac{1}{2})$ with respect to the orthorhombic unit cell. Additional magnetic reflections seen at 4\,K cannot be uniquely indexed nor attributed to any commensurate wave vector. (b) A view of low $2\theta$ Bragg reflections showing emerging peaks in each magnetic phase. (c) The intensity as a function of temperature for magnetic Bragg peaks at 18.6\degree and 9.7\degree marked in panel (b). The solid line represents the background. This shows the onset of a commensurate antiferromagnetic phase (AFM1) with propogation vector \qmagI\ at \TNI. Below \TNII, the incommensurate magnetic order noted at 4\,K in panel (a) emerges.
}
\label{pow_patt}
\end{figure*}
\begin{figure*} \centering \includegraphics [width = \textwidth] {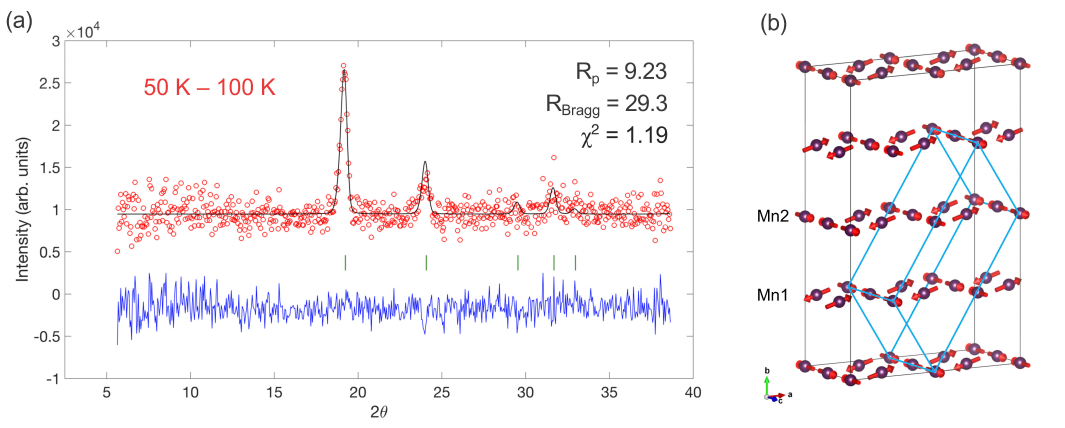}
\caption{Refinement of the magnetic structure of powder neutron diffraction data at 50 K after subtracting 100 K data. (a) Rietveld refinement of the AFM1 phase using a magnetic wavevector \qmagI$= (\frac{1}{2},\frac{1}{2},\frac{1}{2})$ with the difference curve shown in blue. (b) Noncoplanar magnetic structure solution with Mn1 moments in the \textit{ab} plane and Mn2 moments polarized along \textit{c}, with refined moments of $2.43 \mu_{B}$ and $2.5 \mu_{B}$, respectively. The monoclinic magnetic unit cell is marked with the blue parallelepiped.}
\label{magn_rf}
\end{figure*}

Data from neutron diffraction measurements collected at $T =$ 100, 50, and 4\,K using powders ground from single crystals are shown in Figure \ref{pow_patt} (a-b).  Markers at 100 K label several high symmetry nuclear Bragg reflections in Figure \ref{pow_patt} (a). The asterisk near 41\degree\ marks a nuclear peak attributed to residual Bi from the crystal growth flux. At 50\,K, magnetic Bragg reflections appear at low $2\theta$ (low Q) and can be indexed with a reduced wave vector \qmagI $= (\frac{1}{2},\frac{1}{2},\frac{1}{2})$ with respect to the orthorhombic unit cell. Diffraction data collected at a base temperature of 4\,K contain additional magnetic reflections not indexed by \qmagI. These magnetic reflections not indexed by the nuclear unit cell indicate antiferromagnetic (AFM) order.

Figure \ref{pow_patt}(b) shows a zoomed view of the reflections for $2\theta < 31^{\circ}$, which are all resolution limited.
Figure \ref{pow_patt}(c) shows the diffraction intensity versus temperature collected at the two magnetic Bragg peaks marked in Figure \ref{pow_patt}b. The magnetic Bragg reflection intensity is proportional to the square of the magnetic order parameter and tracks the onset and evolution of a new magnetic phase. The onset of intensity at $2\theta = 18.6^{\circ}$ associated with \qmagI\ indicates a N\'{e}el temperature of 72\,K, which is denoted as \TNI. AFM1 denotes the magnetic phase just below this temperature.
Below \TNII\ = 35 K, the peak at 9.7\degree, not indexed by \qmagI, appears. The magnetic phase below \TNII\ if referred to as AFM2, and the data collected at 4\,K show several magnetic Bragg peaks associated with this phase. In the AFM2 state, some magnetic intensity remains at Bragg positions associated with AFM1. There are two possible explanations for this observation: the AFM2 state below may involve the coexistence of the \qmagI\ propagation vector with an additional propagation vector; or, alternatively, a single phase with a unique magnetic propagation vector may account for all the magnetic peaks below \TNII.
However, no commensurate wave vector can index all of the reflections observed in AFM2. Note that other temperatures marked on Figure \ref{pow_patt}c (T$'$ and T*) are defined from physical property measurements discussed below.

Rietveld refinements using the Fullprof Suite \cite{rodriguez1993recent} were performed for the AFM1 phase. The magnetic contribution to the diffraction was isolated by subtracting the data collected in the paramagnetic state (100\,K) from the data collected at 50\,K. Fitting was performed in the range $5^{\circ} < 2\theta < 38.5^{\circ}$ [Figure\ref{magn_rf}a] since no magnetic signal could be resolved above the background at higher angles.

Representational analysis performed using SARAh \cite{wills2000new} for space group $Cmmm$ and propagation vector $(\frac{1}{2},\frac{1}{2},\frac{1}{2})$ yielded two possible irreducible representations,$\Gamma_1$ and $\Gamma_3$ (see Supporting Information).
$\Gamma_1$ constrains the Mn1 moments to the \textit{ab} plane and the moments on Mn2 site to lie along \textit{c}, while in $\Gamma_3$, this relationship is swapped.
The fit using $\Gamma_1$ [Figure\ref{magn_rf}a] gave agreement factors R${_p} = 9.23$, R${_{Bragg}} = 29.3$, and $\chi^{2} = 1.19$, which are defined as the profile R-factor, the magnetic Bragg R-factor, and reduced chi-squared, respectively. The magnetic unit cell, outlined in blue in Figure \ref{magn_rf}b, is monoclinic with magnetic space group $C_a2/m$ (12.64) and cell parameters \textit{a} = 16.33~{\AA}, \textit{b} = 8.26~{\AA}, \textit{c} = 4.16~{\AA}, ${\alpha} = 90^{\circ}$, ${\beta} = 114.05^{\circ}$, and ${\gamma} = 90^{\circ}$ (see Supporting Information). Refinements using $\Gamma_3$ produced the same diffraction pattern and agreement factors, since the two magnetic unit cells are related by a translation of $(\frac{1}{2}, 0, \frac{1}{2})$, leaving the square of the magnetic structure factor invariant. The unpolarized neutron diffraction data cannot distinguish between the two magnetic configurations nor a linear combination of the two irreducible representations.

Figure \ref{magn_rf}b shows the magnetic structure solution in the conventional magnetic cell where the paramagnetic structure is doubled along each of the three crystallographic directions.  Mn2 sites form antiferromagnetic chains with ordered moments of $2.5\ \mu_{B}$ directed along the \textit{c}-axis. The moments on the Mn1 site lie in the \textit{ab} plane, tilted at 16(5)\degree\ from the $a$-axis, with a refined value of $2.43\ \mu_{B}$. This noncoplanar model is favored based on the present analysis; however, constraining the Mn1 moments to lie along the a-axis describes the data nearly as well.

\subsection{Physical properties of \MnRhBi\ single crystals}

The temperature and magnetic field dependence of the physical properties of \MnRhBi\ including effects of the magnetic phase transitions at \TNI\ and \TNII\ were examined using transport, magnetization, and heat capacity measurements. Because of the growth habit and the easy cleavage described above, the mm-scale crystals used for these measurements are best considered as composites of multiple plate-like ribbons that are nearly crystallographically aligned.

\begin{figure} \centering \includegraphics [width = 3.5 in] {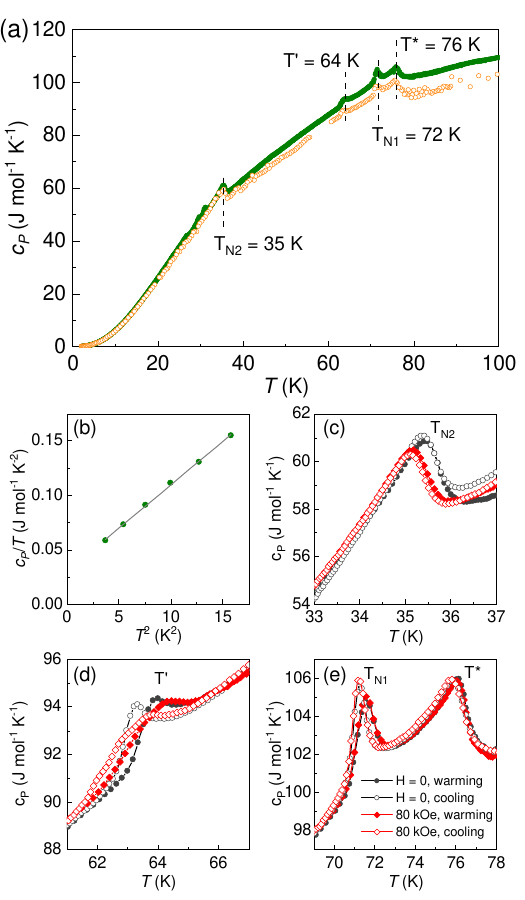}
\caption{
Heat capacity of \MnRhBi.  (a) Results from two samples of single crystal \MnRhBi\ measured in zero magnetic field. Four anomalies are noted by the dashed lines. (b) The low temperature behavior and a linear fit corresponding to $c_P(T) = \gamma T +\beta T^3$. (c-e) Heat capacity near each phase transition determined from analyzing heating and cooling portions of heat pulses in both zero and 80\,kOe applied magnetic field. Here mol denotes moles of formula units. }
\label{xtal-hc}
\end{figure}

Results of heat capacity measurements on \MnRhBi\ crystals are shown in Figure \ref{xtal-hc}a, which includes data from two samples. Note that mol is used here to denote moles of formula units, equivalent to five moles of atoms or one mole of Mn. Figure \ref{xtal-hc}b shows the low temperature data plotted as $c_P/T$ vs $T^2$. Assuming the temperature is far enough below the magnetic ordering transitions such that the heat capacity is due only to free carriers and phonons, the linear fit gives a Sommerfeld coefficient of 30\,mJ/mol/K$^2$ and a Debye temperature of 107\,K.

In both samples, four clear anomalies in the heat capacity are observed. This includes anomalies at both \TNI\ = 72\,K and \TNII\ = 35\,K identified above as antiferromagnetic transitions by neutron diffraction. Additional anomalies are detected just above \TNI\ at T* = 76\,K and below \TNI\ at T$'$ = 64\,K. This indicates the presence of four phase transitions in \MnRhBi\ below room temperature. The higher temperature features coincide with the onset of the anomalous thermal contraction noted above and shown in Figure \ref{structure}. Thus, it is reasonable to tentatively ascribe a structural nature to the transition at T*, although no change in symmetry is detected from the present data between room temperature and 4\,K.

The heat capacity near each phase transition is examined more closely in Figure \ref{xtal-hc}c-e, where results for both heating and cooling in both zero and 80\,kOe applied fields are shown.
The anomaly at T* shows no response to magnetic field, and only a barely detectable thermal hysteresis of about 0.2\,K.
The transition at \TNI\ is significantly sharper but shows a similar behavior; no effect of the field and a thermal hysteresis of about 0.3\,K are seen.
The largest effects are seen at the T$'$ transition, with a thermal hysteresis of about 0.6\,K. The magnetic field strongly broadens this transition, particularly toward lower temperatures.
At \TNII, no thermal hysteresis is detected and a slight suppression in temperature by 0.3\,K is seen in the magnetic field.

Based on the thermal hysteresis, on the transition at T$'$ is likely to have significant first-order character. This transition also has the most pronounced response to magnetic field. The suppression by the field, also seen to a smaller degree for \TNII, is consistent with an antiferromagnetic character. The broadening is unusual and its origin is unclear.

\begin{figure} \centering \includegraphics [width = 3.5in] {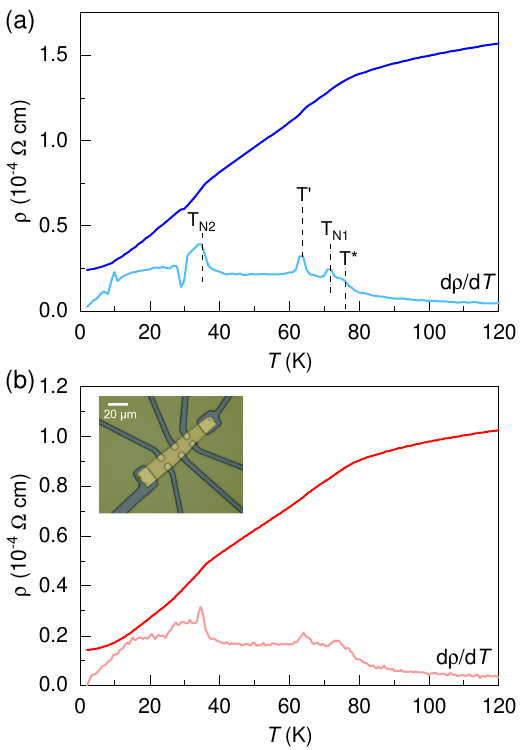}
\caption{
Electrical transport results from single crystal \MnRhBi. The current was applied along the c-axis in all measurements.
(a) Temperature dependence of electrical resistivity of a bulk crystal.
(b) Temperature dependence of electrical resistivity of an exfoliated flake (thickness = 660\,nm).
In both (a) and (b) the temperature derivative is also shown (arbitrary scale), with the main features associated with phase transitions noted in panel (a).
}
\label{xtal-R}
\end{figure}

The electrical resistivity of a bulk, as-grown crystal \MnRhBi\ is shown in Figure \ref{xtal-R}a. The resistivity of an exfoliated sample of thickness 660\,nm in shown in Figure \ref{xtal-R}b. Similar behavior was seen in transport data from a 27\,nm thick sample. The electrical resistivity of the bulk crystal and the thin flake are similar and increase with temperature indicating metal-like behavior. The bulk and exfoliated samples have residual resistivity ratios ($R_{300K}/R_{3K}$) of 8.0 and 8.7, respectively.

Several anomalies are observed in both datasets, and these are highlighted in the temperature derivative curves plotted in Figure \ref{xtal-R}. There is a clear peak in d$\rho$/dT at \TNI, with a shoulder seen at T*. Peaks are also seen at T$'$ and \TNII.
There is also notable structure in the derivative just below the \TNII, which can also be seen in the heat capacity data in Figure \ref{xtal-hc}a.

\begin{figure} \centering \includegraphics [width = 3.4 in] {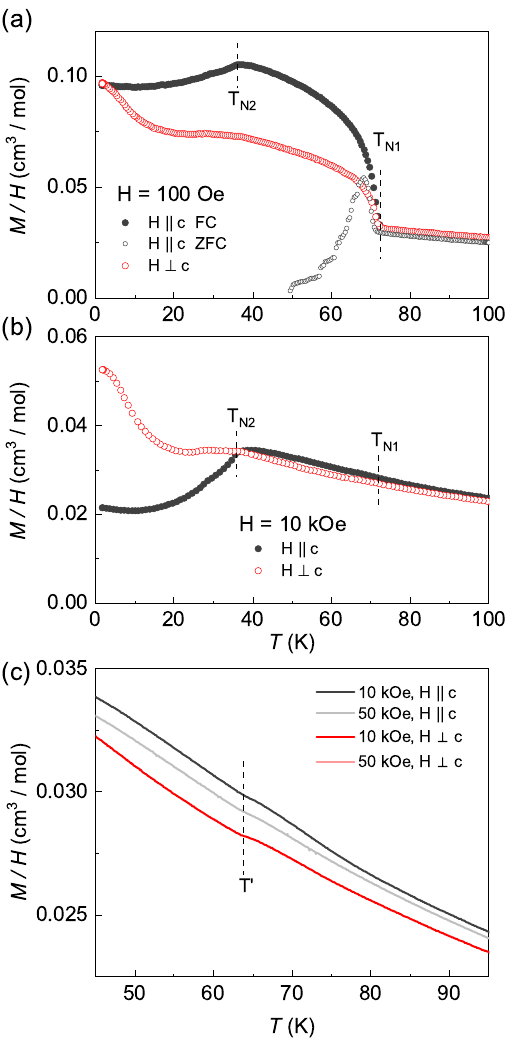}
\caption{
Temperature dependence of the DC magnetic susceptibility of \MnRhBi\ crystals near the phase transition temperatures. Data are shown for fields along the \textit{c}-axis and perpendicular to the \textit{c}-axis. Specifically, the data for $H \perp c$ were collected with the field perpendicular to the long axis and in the natural growth plane of the crystal ($H \perp c$ and $H \perp [110]$).
For both orientations and near \TNI\ and \TNII, low field behavior (100\,Oe) is shown in (a) and intermediate field behavior (10\,kOe) is shown in (b). The magnetic susceptibility in two fields and both orientations near T$'$ is shown in (c).
}
\label{xtal-mag}
\end{figure}

The temperature dependence of the magnetic susceptibility ($M/H$) of \MnRhBi\ is summarized in Figure \ref{xtal-mag}.  The figure includes data from two orientations of the crystals. For $H || c$ the field is along the needlelike growth direction and parallel to the Mn chains. $H \perp c$ corresponds to $H$ in the plane of the platelike crystals, that is, perpendicular to both the c-axis and the [110] direction. In the low field magnetization data (Figure \ref{xtal-mag}a) the two antiferromagnetic transitions identified by neutron diffraction appear prominently. Upon cooling, at \TNI\ the magnetization displays a sharp increase although the value remains small. This is characteristic of a transition into a canted antiferromagnetic state with a small uncompensated moment. Neutron diffraction shows that both Mn sites develop ordered moments at this tempeature. At \TNII\ the magnetization displays a cusp, reminiscent of an antiferromagnetic transition. Since this occurs below \TNI, this can be attributed to a spin reorientation in the ordered state. A strong divergence between field cooled (FC) and zero-field cooled (ZFC) data is seen in Figure \ref{xtal-mag}a. This onsets just below \TNI, and is consistent with the interpretation of a transition into a canted AFM state, where upon cooling in zero (or near zero) field domains can be frozen in and on warming persist up to close to the transition temperature. No notable difference was seen between field-cooled warming and cooling data at 100\,Oe. The fact that the fully compensated AFM1 model fits the 50\,K neutron diffraction data well (Figure \ref{magn_rf}) indicates the canting in this state is small.

At the intermediate field of 10\,kOe (Figure \ref{xtal-mag}b), only the \TNII\ feature is clearly observed. Magnetic anomalies associated with the other transitions are difficult to detect. Figure \ref{xtal-mag}c shows, over a narrow temperature range, the magnetic susceptibility measured at 10 and 50\,kOe. A subtle but visible kink at T$'$ is noted and it does not change significantly between 10 and 50\,kOe. This is likely masked in the low field data (Figure \ref{xtal-mag}a) by the rapid change in \textit{M}(\textit{T}) through this temperature. In addition, a very subtle slope change may be noted in the 72-76\,K  range (Figure \ref{xtal-mag}c), but this cannot be associated with either \TNI\ or T* with any confidence.

If indeed there is no magnetic signature of the phase transition at T*\,=\,76\,K, it may be reasonable to ascribe to it a primarily structural nature. However, since no symmetry change is indicated based on the present data, the details of such a transition remain unclear and further study is warranted. It is likely that this material has substantial magnetoelastic coupling and that may prove to be an interesting direction to explore. Indeed, the heat capacity behavior and likely magneto-structural nature of the transitions at T* and \TNI\ bring to mind the LaFeAsO-family of parent phases of the Fe-based superconductors, in which spin-lattice interactions produce a nematic phase and coupled magnetic and crystallographic phase transitions.\cite{McGuire2008, fernandes2014drives, Kaneko2017} Further work to carefully monitor the structure through the phase transitions in \MnRhBi\ is needed. Note that the structure is orthorhombic at room temperature, so structural distortions along a, b, or c are all allowed without breaking any symmetry (see Figure \ref{structure}d).

\begin{figure} \centering \includegraphics [width = 3.4 in] {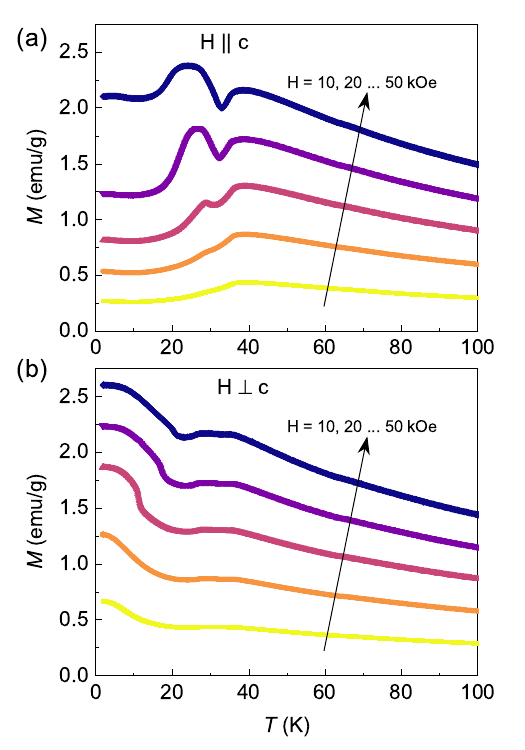}
\caption{
Evolution at of the magnetization of \MnRhBi\ crystals at higher applied fields with data collected upon cooling and in 10\,kOe steps from 10 to 50\,kOe. (a) Data collected with $H || c$. (b) Data collected with $H \perp c$.
}
\label{xtal-magfields}
\end{figure}

As pointed out above, at intermediate and higher field the magnetic response is primarily constrained to temperatures below \TNII. A series of $M$($T$) curves collected upon cooling in fields from 10 to 50\,kOe are shown in Figure \ref{xtal-magfields}. Data are shown for two orientations of the field. The kink at \TNII\ remains present over the entire field range, but the behavior at lower temperature evolves strongly. In particular, a local maximum in $M$ emerges near 24\,K for $H || c$ with associated features appearing in the $H \perp c$ data. This behavior likely points to a complex magnetic ground state and competing magnetic interactions.

It is interesting to consider the interplay of Mn and Rh magnetism as a possible origin of the complicated phase evolution of \MnRhBi. However, neutron diffraction at 50\,K suggests that any ordered moment on Rh is very small in the AFM1 state. DFT calculations performed using a simple antiferromagnetic structure gives a moment on Rh consistent with zero (0.02$\mu_B$). Thus, from the present data it is not possible to identify what role Rh may play in the phase evolution of \MnRhBi.

The magnetic susceptibility measured at 10\,kOe along three orthogonal directions was averaged and used for Curie Weiss fitting. Using data between 100 and 300\,K, the fitting gives an effective moment of 5.9\,$\mu_B$ per formula unit and a Weiss temperature of -94\,K. Similar values were obtained for a polycrystalline sample (see Supporting Information). The effective moment is close to the spin only value expected for Mn$^{2+}$ with a $3d^5$ electronic configuration. The Weiss temperature indicates dominantly antiferromagnetic interactions, which is consistent with the observed magnetic order.

\begin{figure} \centering \includegraphics [width = 3.5in] {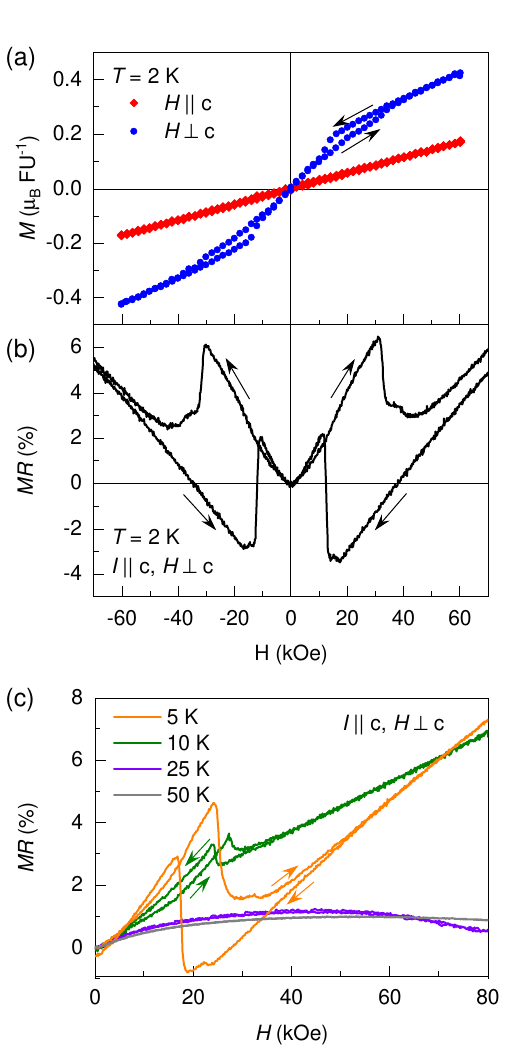}
\caption{
Magnetic field dependence of the magnetization and electrical resistance of \MnRhBi\ crystals. Arrows show the direction the field was swept in different parts of the curves.
(a) Isothermal magnetization data collected at 2\,K showing a field induced transition when the field is applied perpendicular to c and linear behavior when H is along c.
(b) Magnetoresistance (\textit{MR}) with the current (\textit{I}) along the c axis $H \perp c$ showing sharp transitions corresponding to the more subtle features in M(H).
(c) \textit{MR} at higher temperatures showing the field induced transition is still present at 10\,K but is absent at 25\,K.
}
\label{xtal-MH-MR}
\end{figure}

Isothermal magnetization curves measured at 2\,K are shown in Figure \ref{xtal-MH-MR}a. No magnetic saturation is observed up to 60\,kOe, typical of an antiferromagnetic state. With the field along the c-axis, the magnetization is linear between -60 and 60\,kOe, while a field induced transition is apparent when the field is applied perpendicular to c. Recall the $H \perp c$ direction used here is in the plane of the platelike crystals, perpendicular to the [110] direction. For a simple collinear antiferromagnet, one expects to observe linear $M(H)$ when the field is applied perpendicular to the spin direction, and a lower susceptibility at low fields followed by a spin-flop or spin-flip when the field is applied along the spin direction. These expectations are not borne out in the behavior seen in Figure \ref{xtal-MH-MR}a. In addition, the measured moment at the highest field used in the present study reaches only about 0.4\,$\mu_B$ per Mn, much lower than any expected saturation moment. Thus, higher field measurements will be needed to draw any conclusions from the isothermal magnetization curves. Neutron diffraction shows that the magnetic structure is complex and likely incommensurate at low temperature, further complicating the $M(H)$ behavior.

There is field-hysteretic behavior in $M(H)$ for $H \perp c$ at 2\,K. Upon increasing field, a slight increase in $M$ is observed near 32\,kOe. The associated decrease in $M$ upon reducing the field begins near 15\,kOe. Although these features in $M(H)$ are subtle, they are pronounced in the magnetoresistance ($MR$). This is shown in Figure \ref{xtal-MH-MR}b. The critical fields in the field-hysteresis loop in $MR$ agree with those identified in $M(H)$. The field dependence and hysteresis are strong enough to cause a sign change in $MR$, from positive to negative, upon decreasing field before it returns to its low field behavior at the lower critical field. At 2\,K this starts near 15\,kOe and completes sharply near 12\,kOe. The measured magnetoresistance is significant but not particularly large; at 2\,K it reaches a value of 10\% at 120\,kOe.

The sharp changes and hysteretic behavior of \textit{MR} seen at 2\,K persist to higher temperatures, and this is shown in Figure \ref{xtal-MH-MR}c. The magnitude of the changes at the critical fields and the degree of hysteresis are reduced as temperature increases, and they vanish between 10 and 25\,K. At 25 and 50\,K no field induced transitions are seen, and $MR$ is limited to about 1\% and shows concave down field dependence over the field range studied here.

\section{Summary and Conclusions}

Analysis of the crystal structure, calculation of cleavage energy, and demonstration of exfoliation establishes \MnRhBi\ as a van der Waals layered material. Interestingly, the van der Waals gap in this compound resides between neighboring layers of Bi atoms, and the Bi atoms have lone pairs extending into the gap. This may provide new avenues for studies of how electronic properties of the cleavage surface affect interactions and proximity effects in heterostructures. Electrical resistivity measurements show that \MnRhBi\ is metallic, as expected based on the chemical formula, and neutron diffraction measurements show that it adopts a complex and incommensurate antiferromagnetic ground state. Thus, \MnRhBi\ is a somewhat rare example of a cleavable antiferromagnetic metal, a class of compounds expected to have uses in spintronics. In addition to these key behaviors, this compound displays a complex series of temperature and field induced phase transitions with magnetic, electronic, and structural character.

The phase transition at T* = 76\,K is detected in heat capacity and resistivity but not in magnetization, suggesting it may have a structural origin. At \TNI\ = 72\,K, neutron diffraction reveals that antiferromagnetic order with propagation vector ($\frac{1}{2}, \frac{1}{2},\frac{1}{2}$) emerges. This transition is observed in heat capacity, resistivity, and magnetization as well. These measurements also reveal a transition at T$'$ = 64\,K. The heat capacity anomaly at this transition responds most strongly to magnetic field, broadening toward lower temperatures when the field is applied. The nature of this transition is yet unclear, but it must have some magnetic character. Note that the magnetic structure determination reported here was measured at 50\,K, below both \TNI\ and T$'$. Finally, at \TNII\ = 35\,K addition magnetic Bragg peaks emerge, indicating a transition into the incommensurate antiferromagnetic AFM2 ground state, with signatures again seen in heat capacity, resistivity, and magnetization. Magnetization and magnetoresistance measurements suggest that the magnetic structure below \TNII\ evolves in a complex way with applied magnetic field.

This work provides the basic understanding of \MnRhBi\ needed to motivate and inform future studies of this unique material and its potential scientific or technological applications. Several promising directions for further work are identified. Continued crystallographic research should explore magnetoelastic coupling, strain effects, and the structural response at each phase transition. Future work on its physics should be aimed at identifying the magnetic ground state at low temperature, and understanding the roles of spin-orbit coupling on the heavy elements, magnetism on Rh, and lone pair electrons on Bi. Finally, further studies of exfoliated samples are needed, and some are underway, to understand the behavior of this cleavable antiferromagnetic metal in the ultrathin limit.

\section{Methods}

\subsection{Synthesis}

Polycrystalline \MnRhBi\ was synthesized by a long term annealing of the premelted stoichiometric mixture of the elements as reported by Kainzbauer et al.\cite{kainzbauer2018single} \MnRhBi\ single crystals were grown out of Bi flux. The starting materials are elemental Mn pieces (Alfa, 99.99\%), Rh powder (Alfa, 99.9\%), and Bi shots (Alfa, 99.999\%). Mn pieces were first cleaned by subliming away residual oxides after washing with diluted HCl. All starting materials were put in a 2\,ml Canfield-crucible-set \cite{Canfield-2016} at the molar ratio of Mn:Rh:Bi = 1:1:10 and then sealed in a fused silica ampoule under vacuum. The sealed ampoule was heated to 900$\degree$C in 6 hours, homogenized for 16 hours, furnace cooled to 500$\degree$C, and then slowly cooled to 350$\degree$C over a week. At 350$\degree$C, the Bi flux was decanted from the \MnRhBi\ single crystals.

\subsection{Experimental measurements and characterization of powders and crystals}

X-ray powder diffraction measurements were performed on a PANalytical X’Pert Pro MPD with a Cu-K$_{\alpha,1}$ incident-beam monochromator for phase identification determination of the lattice parameters. Low temperature diffraction data were collected using an Oxford PheniX closed cycle cryostat. The refinement of all powder diffraction patterns in this work was performed using FullProf. Polycrystalline samples were nearly single phase, with only a few percent of Bi or \ce{Mn5Rh6Bi18} as secondary phases.

Elemental analysis was carried out on as-grown crystals using a Hitachi-TM3000 microscope equipped with a Bruker Quantax 70 EDS system. The average measured atomic ratio was Mn:Rh:Bi = 21:23:56. This is close to the expected ratio of 1:1:3 considering the semiquantitative nature of the technique and the significant overlap between Rh-L and Bi-M lines.

Neutron powder diffraction was performed on the HB-2A beamline at the High Flux Isotope Reactor (HFIR) at Oak Ridge National Laboratory (ORNL). Approximately 0.3 g of crystals were ground into a fine powder, placed into a vanadium sample can 3 mm in diameter to reduce the absorption effect of Rh. Diffraction patterns were collected from 100 K down to a base temperate of 4 K, with collimator settings open-open-12’, and a Ge(113) monochromator provided an incident wavelength of $\lambda = 2.41$ \AA. The patterns were collected over a Q-range of $0.45 $ \AA$^{-1} < Q < 4.19 $ \AA$^{-1} (5^\circ < 2\theta <125.5^\circ)$ with count times of 8 hours per scan.

Magnetization, heat capacity, and transport properties were measured using standard practices with commercial cryostats from Quantum Design: Magnetic Property Measurement Systems (MPMS-XL and MPMS3) and Physical Property Measurement Systems (PPMS).

\subsection{Density functional theory calculations}

All theoretical studies were conducted using density functional theory (DFT) with the projector augmented plane-wave (PAW) method, as implemented in VASP version 6.3.2~\cite{Kresse1994,PBE1996}. PAW potentials for Mn (3p$^6$4s$^2$3d$^5$), Rh (4d$^8$5s$^1$), and Bi (5d$^{10}$6s$^2$6p$^3$) were employed. All calculations were performed with an energy cutoff set at 800 eV, and Monkhorst-Pack k-point grids of 4$\times$2$\times$8 for the bulk structure and 4$\times$1$\times$8 for the slab structure. Ionic positions were relaxed until forces on each atom were reduced below 10 meV/\AA, achieving a total energy convergence criterion of 10$^{-6}$ eV. To accurately capture the van der Waals (vdW) interactions within the MnRhBi$_3$ structure, the van der Waals density functional (vdW-DF)~\cite{vd_DF2004, Thonhauser-2007} with the optB86b exchange functional was employed~\cite{MK2011}. The crystal structure of the bulk and slab MnRhBi$_3$ models are shown in the Supporting Information. The vdW-DF-optB86b relaxed lattice constants are $a$ = 8.755 \AA, $b$ = 13.775 \AA, and $c$ = 4.144 \AA. The distance between Bi-Bi layers ($d_0$) is 3.60 \AA, indicating the separation between these layers within the structure, and are in good agreement with the experimental interlayer distance.

\subsection{Exfoliation and characterization of exfoliated samples}

Samples were exfoliated with standard scotch tape techniques on Si substrates covered with 285 nm thick thermally grown \ce{SiO2}. Before exfoliation Si/\ce{SiO2} substrates were treated with \ce{O2} plasma for 10 minutes at the RF power of 90 W. Optical microscopy was used to identify flakes suitable for device fabrication and selected flakes were investigated using atomic force microscopy. Suitable thin flakes were then covered with PMMA (Polymethylmethacrylate) and standard electron beam lithography, thermal evaporation of Cr/Au stack and liftoff were used to fabricate devices, an example of one of such devices is shown in Figure \ref{xtal-R}b. To protect thin flakes of \MnRhBi\ from possible degradation we performed all fabrication and characterization inside an Ar-filled glovebox. Before performing transport measurements, samples were covered with another layer of PMMA for protection. \cite{Ovchinnikov2021}

Transport measurements were performed with standard lock-in techniques using the Physical Property Measurement System (PPMS, Quantum Design, 2 K – 300 K, 9 T). Currents in the range of 1 – 400 $\mu$A depending on the sample thickness and dimensions were applied to the drain and source electrodes of samples in the shape of Hall bars. We used lock-in amplifiers (Stanford Research Systems SR830 and SR860) to source current and low-noise current amplifiers (DL 1210) and low-noise voltage amplifiers (Stanford Research Systems SR560) in tandem with lock-in amplifiers to measure current and voltage. All measurements used low frequencies (between 9.999 Hz and 17.777 Hz) 

Atomic force measurements on exfoliated flakes were performed inside an Ar-filled glovebox using Bruker Dimension Icon AFM after exfoliation and before further device fabrication. The peak force mode available in Bruker Icon was mainly used, while contact and tapping AFM modes provided consistent results.

\section{Supporting Information}

Rietveld fit to neutron powder diffraction data at 100\,K, representation analysis results and magnetic space group, properties of polycrystalline \MnRhBi, exfoliation of ultrathin samples, electronic structure and cleavage energy calculations (PDF).

\section{Acknowledgments}

Work was primarily supported by the U.S. Department of Energy, Office of Science, Basic Energy Sciences, Division of Materials Sciences and Engineering at Oak Ridge National Laboratory. This research used resources at the High Flux Isotope Reactor, a U.S. DOE Office of Science User Facility operated by the Oak Ridge National Laboratory.
For exfoliation and transport measurements on exfoliated flakes, D.O. acknowledges the KU Research Go Award (1004170) and KU Startup Funding and X.X. acknowledges support from DOE BES (DE-SC0012509).

\section{Supporting Information for \\ MnRhBi$_3$: A Cleavable Antiferromagnetic Metal}

\subsection{Rietveld refinement of neutron powder diffraction data}

\begin{figure} \centering \includegraphics [width = 4.0in] {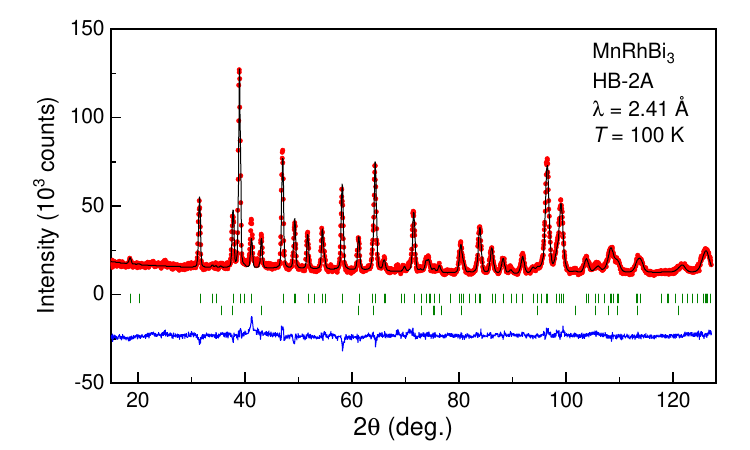}
\caption{
Data points collected at 100\,K on HB-2A along with the Rietveld fit (black line) and difference curve (blue line). Upper tick marks locate reflections from the reported crystal structure. Lower tick marks locate reflection from a residual Bi impurity of about 7\% by weight.
}
\label{Riet}
\end{figure}

Figure \ref{Riet} shows the Rietveld refinement of data collected at 100\,K giving the crystal structure parameters listed in the main text. Agreement factors are Rp = 5.24, Rwp = 6.65, $\chi^2$ = 2.27.

\subsection{Magnetic structure: Representation analysis and magnetic space group.}

Representation analysis results from SARAh \cite{wills2000new} for space group $Cmmm$ and propagation vector $(\frac{1}{2},\frac{1}{2},\frac{1}{2})$ are listed in Table \ref{sarah}. Magnetic space group information is given in Table \ref{msg}.

\begin{table}
\caption{\label{sarah}Basis vectors for the space group $Cmmm$ with $\textbf{k}_{12} = (\frac{1}{2}, \frac{1}{2}, \frac{1}{2})$. The decomposition of the magnetic representation for the Mn site $(0, 0, 0)$ is $\Gamma_{Mag} = 2\Gamma^{1}_{1} +1\Gamma^{1}_{3}$. The atoms of the nonprimitive basis are defined according to $\textrm{Mn}1: (\frac{1}{2}, 0, \frac{1}{2}), \textrm{Mn}2: (0, 0, 0)$.}
\begin{tabular}{ccccccccc}
\hline
 \multicolumn{1}{c}{\textrm{IR}} & \multicolumn{1}{c}{\textrm{BV}} & \multicolumn{1}{c}{\textrm{Atom}} & \multicolumn{6}{c}{\textrm{BV components}}\\
  & & & $m_a$ & $m_b$ & $m_c$ & $im_a$ & $im_b$ & $im_c$\\ \hline
 $\Gamma_1$ & $\psi_1$ & 1 & 4 & 0 & 0 & 0 & 0 & 0 \\
         &       & 2 & 0 & 0 & 4 & 0 & 0 & 0 \\
         & $\psi_2$ & 1 & 0 & 4 & 0 & 0 & 0 & 0 \\
         \\
 $\Gamma_3$ & $\psi_3$ & 1 & 0 & 0 & 4 & 0 & 0 & 0 \\
         &       & 2 & 4 & 0 & 0 & 0 & 0 & 0 \\
         & $\psi_4$ & 2 & 0 & 4 & 0 & 0 & 0 & 0 \\
\hline
\end{tabular}
\end{table}
\begin{table}
\caption{\label{msg} Magnetic structure of \MnRhBi\ described under its MSG (crystallographic description), with basic information about its relation with its parent paramagnetic structure. All items in the table are supported by the magnetic CIF format (magCIF).}
\begin{tabular}{lc}
\hline
Compound	&	MnRhBi3	\\	
Parent space group	&	Cmmm (N. 65)	\\	
Propagation vector(s)	&	 (½, ½, ½ )	\\	
Transformation from parent	&	(2a,2b,2c;0,0,0)	\\	
basis to the one used	&		\\	
MSG symbol	&	Ca2/m 	\\	
MSG number	&	12.64	\\	
Transformation from basis used	&	(1/2a+1/2b,c,1/4a-1/4b;0,0,0)	\\	
to standard setting of MSG	&		\\	
Magnetic point group	&	2/m1 (5.2.13) 	\\	
Unit cell parameters (Å)	&	a=17.68       a=90º	\\	
	&	b=27.31       b =90º	\\	
	&	c=8.23          g =90º	\\	\hline
MSG symmetry operations	&	x,y,z,+1	\\	
	&	-x,-y,z,+1	\\	
	&	-x,-y,-z,+1	\\	
	&	 x,y,-z,+1	\\	\hline
MSG symmetry centering operations	&	x,y,z,+1	\\	
	&	x,y+1/2,z+1/2,+1	\\	
	&	x+1/4,y+1/4,z+1/2,+1	\\	
	&	x+1/4,y+3/4,z,+1	\\	
	&	x+1/2,y,z+1/2,+1	\\	
	&	x+1/2,y+1/2,z,+1	\\	
	&	x+3/4,y+1/4,z,+1	\\	
	&	x+3/4,y+3/4,z+1/2,+1	\\	
	&	x,y,z+1/2,-1	\\	
	&	x,y+1/2,z,-1	\\	
	&	x+1/4,y+1/4,z,-1	\\	
	&	x+1/4,y+3/4,z+1/2,-1	\\	
	&	x+1/2,y,z,-1	\\	
	&	x+1/2,y+1/2,z+1/2,-1	\\	
	&	x+3/4,y+1/4,z+1/2,-1	\\	
	&	x+3/4,y+3/4,z,-1	\\	\hline
Positions of magnetic atoms	&	Mn1 Mn 0.25000 0.00000 0.25000	\\	
	&	Mn2 Mn 0.00000 0.00000 0.00000	\\	\hline
Positions of non-magnetic atoms	&	Bi1 Bi 0.00000 0.16890 0.00000	\\	
	&	Bi2$_1$ Bi 0.09715 0.06220 0.25000	\\	
	&	Bi2$_2$ Bi 0.90285 0.06220 0.25000	\\	
	&	Mn1 Mn 0.25000 0.00000 0.25000	\\	
	&	Mn2 Mn 0.00000 0.00000 0.00000	\\	
	&	Rh1 Rh 0.15060 0.00000 0.00000	\\	\hline
\end{tabular}
\end{table}

\subsection{Physical properties of polycrystalline samples}

\begin{figure} \centering \includegraphics [width = 3.0in] {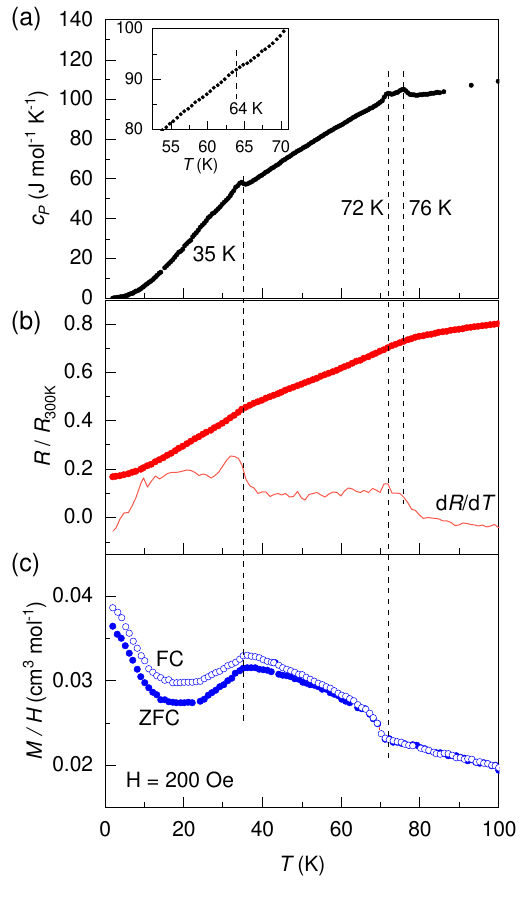}
\caption{
Summary of physical properties of polycrystalline \MnRhBi:
(a) Specific heat capacity. The inset shows data near T$'$.
(b) Electrical resistivity normalized to its value at 300\,K, with the derivative $d\rho/dT$ also shown (arbitratry scale)
(c) Field cooled (FC) and zero field cooled (ZFC) DC magnetic susceptibility (M/H) measured on warming in a 200\,Oe magnetic field.
Dashed lines indicate anomalies associated with phase transitions.
}
\label{polysum}
\end{figure}

Heat capacity, resistivity, and magnetization measurements were performed on polycrystalline pellets made from \MnRhBi\ powders for comparison with the single crystal results. The results are summarized in  Fig. \ref{polysum}.  The phase transitions at T*, \TNI, and \TNII\ can be clearly seen in one or more of the measured properties, manifested as anomalies that are very similar to those seen in the single crystal data. In the polycrystalline samples, T$'$ is difficult to detect, but the indication of a feature near 64\,K may be just visible in the heat capacity as shown in the inset of Fig. \ref{polysum}a.

\begin{figure} \centering \includegraphics [width = 3.0in] {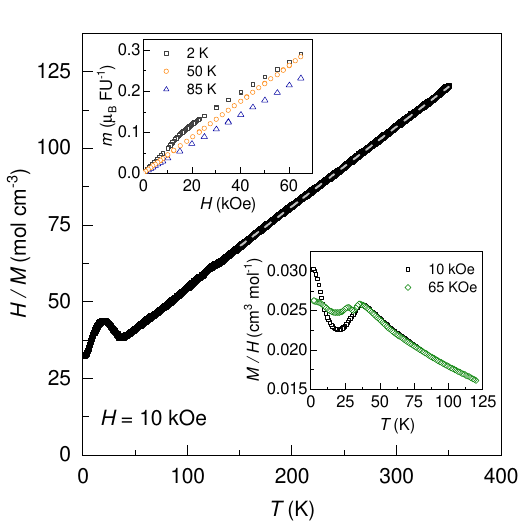}
\caption{
Magnetic behavior of polycrystalline \MnRhBi. The main panel shows the inverse susceptibility (H/M) with the dashed line corresponding to a linear fit between 150 and 350\,K.
The lower inset shows the low temperature behavior at 10 and 65\,kOe.
The upper inset shows isothermal magnetization curves at three temperatures, revealing field induced changes at 2\,K with linear behavior at higher T.
}
\label{polymag}
\end{figure}

The magnetic behavior of the polycrystalline material at higher magnetic fields is shown in Fig. \ref{polymag}. At low temperature, field induced transitions similar to those seen in the single crystals are apparent in the two insets. The main panel shows a Curie Weiss fit to data collected at 10\,kOe between 150 and 350\,K. The fitted effective moment and Weiss temperature are 5.5\,$\mu_B$ and -113\,K, respectively. These values are similar to those determined from the average single crystal magnetic susceptibility given in the main text (5.9\,$\mu_B$ and -94\,K).

\subsection{Exfoliation of ultrathin specimens}

\begin{figure} \centering \includegraphics [width = 4.0in] {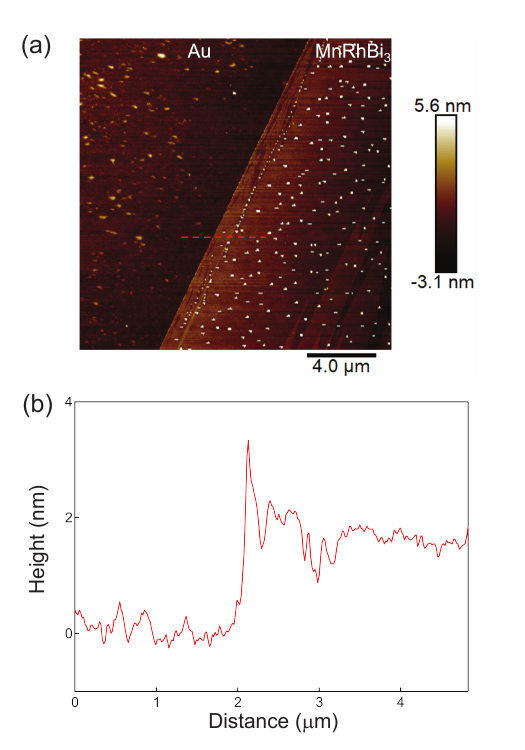}
\caption{
(a) Atomic force microscopy image of an ultrathin \MnRhBi\ flake after exfoliation onto an Au substrate. (b) Line cut from the image showing the flake has a thickness of about 2\,nm.
}
\label{2nm}
\end{figure}

To realize ultrathin flakes of \MnRhBi, exfoliation  was performed on Si/\ce{SiO2} substrates covered with a $\sim$2 nm layer of Cr or Ti followed by a $\sim$3 nm layer of Au. This process \cite{Velicky2018} allowed achieving few nm thick flakes of \MnRhBi. An example is shown in Figure \ref{2nm}, with atomic force microscopy data showing a thickness of 2\,nm. Since the layer spacing is \textit{b}/2 = 0.68\,nm, the measured thickness corresponds to three \MnRhBi\ layers or 1.5 conventional unit cells (see main text Figure 1).

\subsection{Electronic structure}

\begin{figure}
\centering
\includegraphics[scale=0.4]{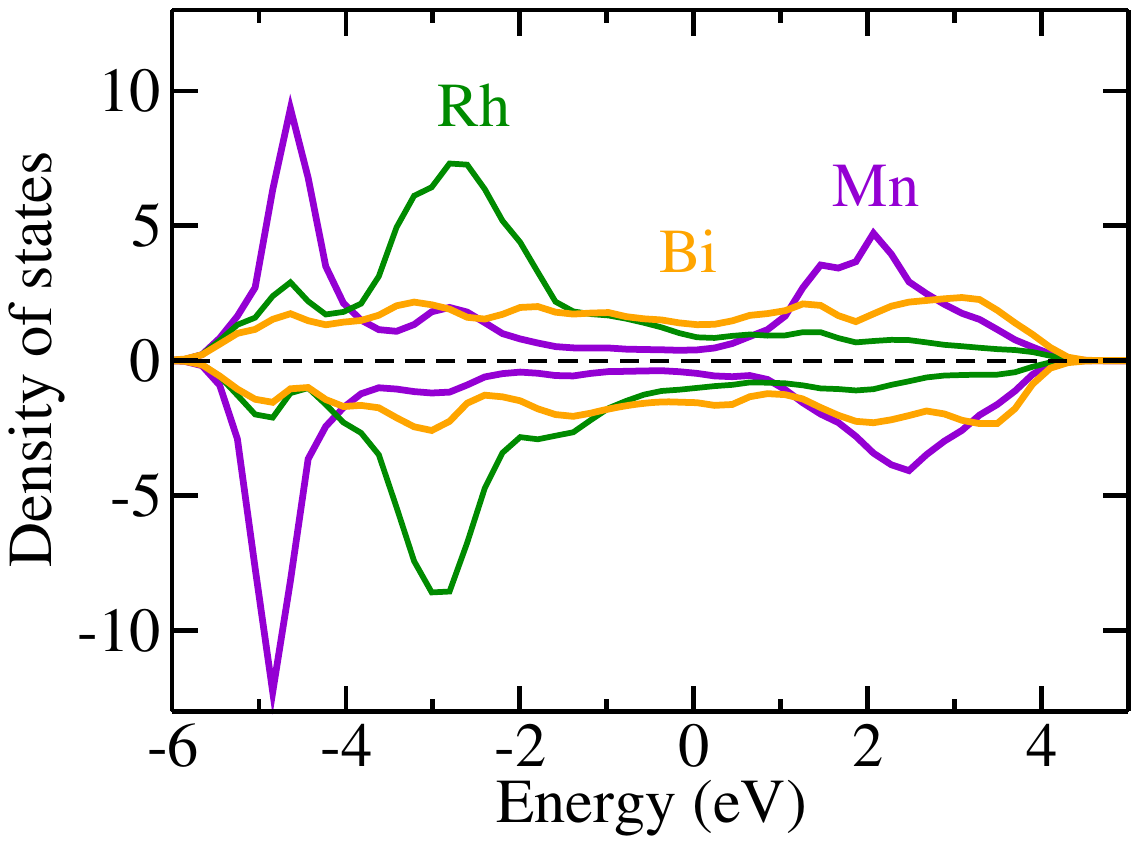}
\caption{Density of states of AFM-MnRhBi$_3$}
\label{fig:dos}
\end{figure}

The vdW-DF-optB86b optimized structures were used to compare the energies of different magnetic phases. Since the experimental magnetic ground state is complex (likely non-collinear and incommensurate), a simple AFM model was constructed and compared to FM ordering. The AFM model used has Mn spins that are antiparallel to nearest Mn atoms both in-plane and out-of-plane. It was found to be 0.5 meV/f.u. lower in energy than the FM state. Calculated magnetic moments are ~4.4 $\mu_B$/Mn, ~0.17 $\mu_B$/Rh and ~0.18 $\mu_B$/Bi. The calculated density of states, depicted in Fig.~\ref{fig:dos}, exhibits the expected metallic behavior for the antiferromagnetic phases. At the Fermi level, the contribution from Bi is the most significant, followed by Rh, with Mn contributing the least.

\subsection{Cleavage energy calculations}

\begin{figure}
\centering
\includegraphics[width = 4 in]{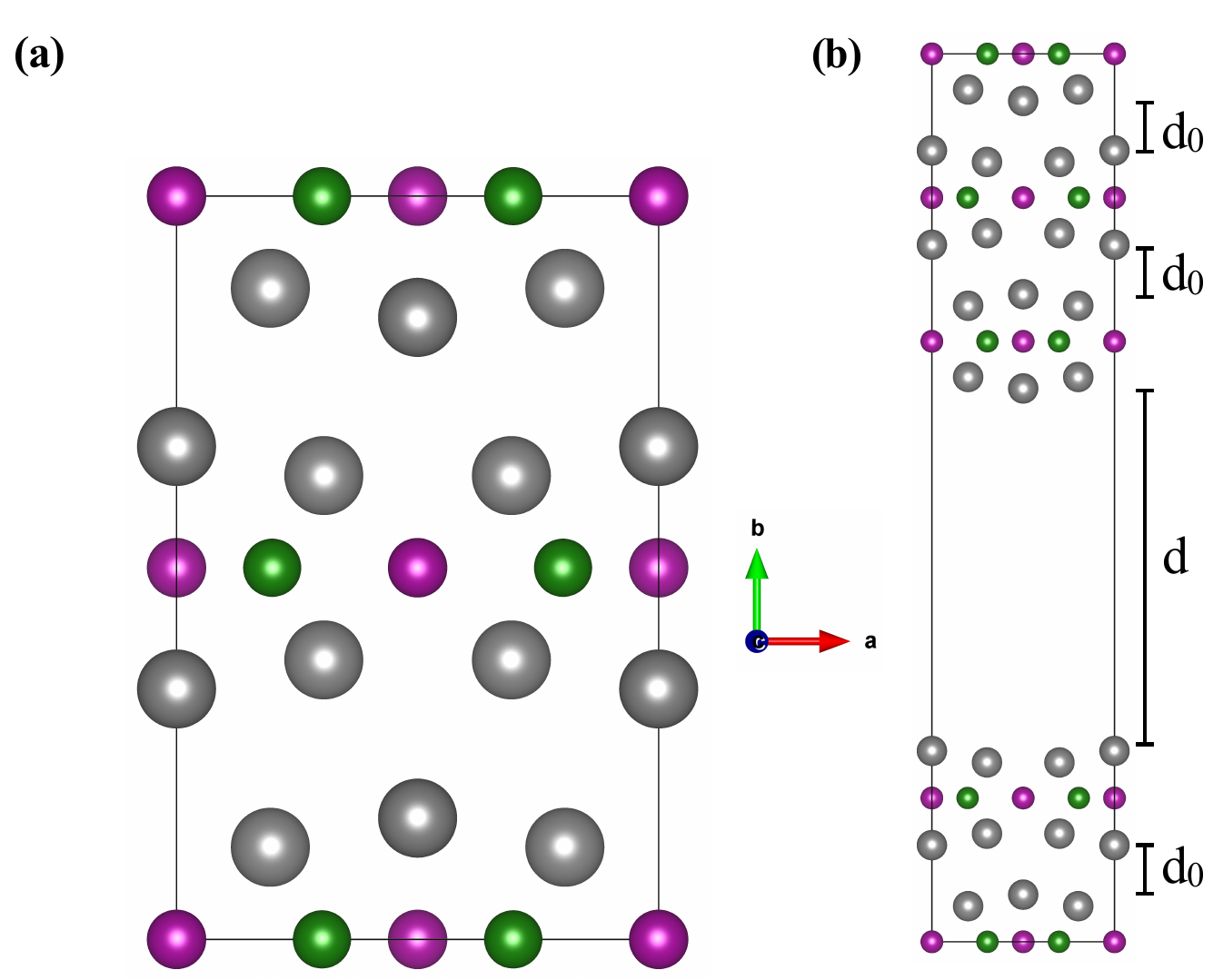}
\caption{(a) Bull crystal structure and (b) slab model depicting layers separated by the equilibrium distance (d$_0$) between Bi-Bi atoms, with a distance (d) of 18\,{\AA} between the top and bottom of the slab along the b axis of MnRhBi$_3$. Mn atoms are shown in violet, Rh atoms in green and Bi atoms in orange.}
\label{fig:slab}
\end{figure}

The cleavage energy ($E_{c}$) is crucial for understanding the material's structural integrity and interlayer bonding characteristics. To calculate the energy required to separate the layers, a vacuum space of 18 \AA~was introduced to minimize interactions between the top and bottom of the slab in the slab structural model (see Fig.~\ref{fig:slab} (b)). The cleavage energy, $E_{c}$, was computed as:
\begin{equation}
E_{c}=\frac{E_{\mathrm{slab}}-E_{\mathrm{bulk}} \frac{N_{\mathrm{slab}}}{N_{\mathrm{bulk}}}}{A_{\mathrm{slab}}}
\label{eq:ec}
\end{equation}
\\
Here, $E_{\mathrm{bulk}}$ is the energy of the bulk system, and $E_{\mathrm{slab}}$ is the energy of the optimized slab configuration after relaxing the atomic positions. $N_{\mathrm{bulk}}$ and $N_{\mathrm{slab}}$ represent the number of atoms in the bulk and slab, respectively, and $A_{\mathrm{slab}}$ denotes the area of the top of the slab. The calculated $E_c$ of MnRhBi$_3$ is 0.56 J/m$^2$.

\bibliography{references_MnRhBi3.bib}

\end{document}